\documentclass[a4paper,10pt]{article}
\newcommand{\nin}{\noindent}
\newcommand{\be}{\begin{equation}}
\newcommand{\ee}{\end{equation}}
\newcommand{\bea}{\begin{eqnarray}}
\newcommand{\eea}{\end{eqnarray}}
\newcommand{\br}{\hskip .25cm/\hskip -.25cm}
\newcommand{\hf}{\frac{1}{2}}
\newcommand{\nn}{\nonumber\\}

\newcommand{\ol}{\overline}

\begin{document}

\begin{center}
{\Large{\bf Path integral quantization of scalar fluctuations\\
 above a kink}}

\vspace{1cm}

{\bf J. Alexandre}$^{1}$ and {\bf K. Farakos}$^{2}$

\vspace{0.5cm}

{\it
1 Department of Physics, King's College London, WC2R 2LS, UK.
 
2 Department of Physics, National Technical University of Athens,\\
Zografou Campus, 15780 Athens, Greece}

\vspace{1cm}

{\bf Abstract}

\end{center}

\vspace{0.5cm}

We quantize scalar fluctuations in 1+1 dimensions above a classical background kink.
The properties of the effective action for the corresponding classical field are studied with an 
exact functional method, alternative to exact Wilsonian renormalization, where the 
running parameter is a bare mass, and the regulator of the quantum theory is fixed.
We extend this approach, in an appendix, to a Yukawa interaction in higher dimensions.

\vspace{1cm}

\section{Introduction}

In the context of higher dimensional field theories, topological defects have been used to explain localization 
of matter on four-dimensional branes. If one considers a scalar field defined on a non trivial vaccum, with
the shape of a kink centered on the brane, it is well known that massless chiral fermions coupled to the scalar field
are localized on the corresponding brane \cite{local}.
 In addition, this localization procedure can be used to define chiral fermions on the lattice, by 
using a kink-like mass term in the extra dimension \cite{lattice1}.
Similar non trivial topological effects, including sphalerons in real time simulations,
also lead to the description of chiral fermions on the lattice \cite{latticefarakos}.

Analytical arguments towards this localization process are given at a classical level, 
and we consider here the quantization of
scalar fluctuations above the kink, in order to exhibit non-perturbative properties. This quantization 
is stable in 1+1 dimensions only, if we consider the scalar field alone \cite{jackiw}, what we will 
do here, since the corresponding toy model exhibits the main features we are interested in. 
 In order to explain more specifically our motivations, though, we
set up, in Appendix B, the first steps of the generalization to a d+1 dimensional Yukawa model, where fermions 
interact with scalar fluctuations above the kink.
The present treatment does not take into account the collective coordinate corresponding to the
translation invariance of the kink \cite{jackiw}, since, in the spirit of the above mentioned 
fermion localization problem, we consider here quantum fluctuations above one specific kink only,
centered on $z=0$, and we do not quantize the whole scalar theory, which contains a degenerate
family of kinks. This is done in different papers \cite{kink1+1}, using canonical quantization.
In more than 1+1 dimension, stability of quantum fluctuations necessitate the presence of another
field than the scalar, and, in this context, the
BRST quantization of the non-linear O(3) model was studied in \cite{kink2+1}.

The method we use here is an alternative to exact Wilsonian renormalization \cite{Wilson}, where, instead of
having fixed bare parameters and a running cut off, we keep a fixed cut off and consider a running bare mass, 
in the spirit of ``functional Callan Symanzik equations'' \cite{initial}. 
The non-perturbative feature of this method, together with the absence of a 
running cut off, led to the derivation of a cut-off-independent dynamical mass generated in the 
framework of a Kaluza-Klein model \cite{kaluzaklein}. This method was also used 
for the description of time-dependent bosonic string actions \cite{string}, where a world sheet cut
off needed to be avoided, and where this alternative approach leads to new results, 
by studying the evolution of the quantum theory with the amplitude of the string tension.

In the present work, the cut off which regulates the evolution equation for the effective theory 
will not appear in the evolution of the dressed parameters, and the 
logarithmic divergences expected in 1+1 dimensions are absent from our flows in the bare mass. Indeed, 
these flows are obtained after a differentiation 
with respect to the bare mass, which is equivalent to inserting an additional propagator 
in the graphs and therefore has the effect of reducing their degree of divergence.
The physical interpretation of the evolution of the quantum theory with a bare mass 
is to control the amplitude of quantum fluctuations: when the bare mass is large,
quantum fluctuations are frozen and the system is almost classical. As the bare mass
decreases, quantum fluctuations gradually appear in the system, which therefore becomes dressed.
A review can be found in \cite{review}.

From a technical point of view, this method can be seen as a tool, used to investigate properties of the quantum theory:
the evolution in the bare parameter leads to a functional partial differential equation, which is then 
split into a series of differential equations, involving the dressed parameters 
which describe the effective theory. The integration of these non-perturbative differential equations leads 
to the effective theory, which exhibits the quantum properties of the system.

\vspace{0.5cm}

Section 2 describes the scalar model we study here, and shows the derivation of the evolution equation for the 
quantum theory with the bare mass of the quantum field which 
fluctuates above a classical background kink. 
The evolution equation we arrive at technically looks like an exact Wilsonian renormalization
equation, but is actually very different in essence, as explained above. 

Section 3 derives the evolution of the dressed parameters defining the quantum system, and
discusses different properties of quantum theory. We show there that no odd power of the 
classical field is present in the effective action, whereas a cubic interaction is present in the bare action.
We also compare our results to one loop predictions, and give new relations on the dressed parameters,
beyond one-loop, as a consequence of the ressumation provided by our evolution equations.

Finally, section 4 contains a general discussion on our results, based on symmetry properties of the quantum 
theory. Appendix A shows the derivation of the evolution equations and Appendix B displays the first steps
on how to generalize the method to a d+1 dimensional Yukawa model.

\section{Model and evolution of the effective theory}

The bare action in 1+1 dimensions is 
\be
S_0=\int dt dz\left\{\hf\partial_\mu\Phi\partial^\mu\Phi-U_B(\Phi)\right\}
\ee
where $z$ is the space coordinate, and the bare potential $U_B(\Phi)$
implements a spontaneous symmetry breaking:
\be\label{barepot}
U_B(\Phi)=-\frac{m_0^2}{2}\Phi^2+\frac{\lambda_0}{24}\Phi^4.
\ee
In 1+1 dimensions, the scalar field has mass dimension 0, which leads to an important 
renormalization property: all the powers of the field are (classically) relevant operators,
and all the coupling constants have mass dimension 2. As a consequence, the bare potential 
(\ref{barepot}) is not chosen on the basis of relevance/irrelevance of the interactions, but 
rather on the assumption of small amplitude of fluctuations above the kink. This assumption will 
prove to be valid, what will be seen with the effective theory that is obtained.\\
The classical equation of motion for the field is
\be\label{eq.mot.}
\partial_\mu\partial^\mu\Phi+U_B^{'}(\Phi)=0,
\ee
where a prime denotes a derivative with respect to $\Phi$. 
We concentrate on the kink solution of eq.(\ref{eq.mot.}) which depends on $z$ only and reads
\be\label{kink}
\Phi_{bg}(z)=m_0\sqrt\frac{6}{\lambda_0}\tanh(\zeta).
\ee
where the dimensionless coordinate $\zeta$ is defined as
\be
\zeta=\frac{m_0z}{\sqrt 2}.
\ee
We consider then the quantum fluctuations $\tilde\Phi$ around $\Phi_{bg}$ and write 
\be
\Phi(t,z)=\Phi_{bg}(z)+\tilde\Phi(t,z).
\ee
If we take into account the equation of motion (\ref{eq.mot.}), the action depending on the 
dynamical variable $\tilde\Phi$ is
\bea
S&=&\int dt dz\Bigg\{\hf\partial_\mu\tilde\Phi\partial^\mu\tilde\Phi
-m_0^2\tilde\Phi^2-\frac{\lambda_0}{24}\tilde\Phi^4\nn
&&~~~~~~~~~~~+\frac{3}{2}m_0^2\left[1-\tanh^2(\zeta)\right]\tilde\Phi^2
-m_0\sqrt\frac{\lambda_0}{6}\tanh(\zeta)\tilde\Phi^3\Bigg\},
\eea
We are interested in studying the quantum theory on the kink background,
and we will derive for this the evolution of the effective action with the bare mass $m_0$.
We will therefore start with the following bare action
\bea\label{bareaction}
S_\xi&=&\int dt dz\Bigg\{\hf\partial_\mu\tilde\Phi\partial^\mu\tilde\Phi
-\xi m_0^2\tilde\Phi^2-\frac{\lambda_0}{24}\tilde\Phi^4\nn
&&~~~~~~~~~~~+\frac{3}{2}m_0^2\left[1-\tanh^2(\zeta)\right]\tilde\Phi^2
-\frac{g_0}{6}\tanh(\zeta)\tilde\Phi^3\Bigg\},
\eea
where the dimensionless parameter $\xi$ controls the amplitude of the mass term $m_0^2\tilde\Phi^2$,
and $g_0=m_0\sqrt{6\lambda_0}$.
We will show that it is possible to derive an exact evolution equation for the effective action with $\xi$.
The corresponding flows describe the evolution from $\xi>>1$, where 
the mass term dominates the Lagrangian and the theory is almost classical,
to the expected quantum theory, obtained for $\xi=1$. 

\vspace{0.5cm}

We now proceed to the quantization of the system, integrating over the dynamical field $\tilde\Phi$.
The partition function is 
\bea
Z_\xi&=&\int{\cal D}[\tilde\Phi]\exp\left(iS_\xi[\tilde\Phi]
+i\int dt dz~j\tilde\Phi\right)\nn
&=&\exp\left(iW_\xi[j]\right),
\eea
where $j$ is the source and $W_\xi$ is the connected graphs generator functional.
The functional derivative of the latter defines the classical field $\phi$:
\bea\label{derivativesW}
\frac{\delta W_\xi}{\delta j}&=&\left<\tilde\Phi\right>_\xi=\phi_\xi\nn
\frac{\delta^2 W_\xi}{\delta j\delta j}&=&-i\phi_\xi\phi_\xi+i\left<\tilde\Phi\tilde\Phi\right>_\xi\nonumber,
\eea
where
\be
\left<\cdot\cdot\cdot\right>_\xi=\frac{1}{Z_\xi}\int{\cal D}[\tilde\Phi](\cdot\cdot\cdot)
\exp\left(iS_\xi+i\int dt dz~j\tilde\Phi\right).
\ee
The effective action $\Gamma_\xi$ (the proper graphs generator functional) is defined as the Legendre transform of $W_\xi$:
after inverting the relation $j\to\phi_\xi$ to $\phi\to j_\xi$, one writes
\be
\Gamma_\xi=W_\xi-\int dt dz~j_\xi\phi,
\ee
where the source $j_\xi$ has now to be seen as a functional of $\phi$, parametrized by $\xi$.
The functional derivatives of $\Gamma$ are then:
\bea\label{derivativesG}
\frac{\delta\Gamma_\xi}{\delta\phi}&=&-j_\xi\nn
\frac{\delta^2\Gamma_\xi}{\delta\phi\delta\phi}&=&-\frac{\delta j_\xi}{\delta\phi}=
-\left(\delta^2 W_\xi\right)^{-1}_{jj},\nonumber
\eea
The evolution equation for $W_\xi$ with the parameter $\xi$ is
\bea
\dot W_\xi&=&-m_0^2\int dt dz~ \left<\tilde\Phi^2\right>\nn
&=&-m_0^2\int dt dz~\phi^2+im_0^2\mbox{Tr}\left\{\frac{\delta^2 W_\xi}{\delta j\delta j}\right\},
\eea
where a dot over a letter represents a derivative with respect to $\xi$.
For the evolution of the effective action $\Gamma$, one should remember that its independent variables are 
$\xi,\phi$, such that
\be
\dot\Gamma_\xi=\dot W_\xi+\int dt dz\frac{\delta W_\xi}{\delta j}\partial_\xi j-\int dt dz~\partial_\xi j\phi
=\dot W_\xi.
\ee
Using the previous results, we finally obtain
\be\label{evolG}
\dot\Gamma_\xi+m_0^2\int dt dz~\phi^2
=-im_0^2\mbox{Tr}\left\{\left(\frac{\delta^2\Gamma_\xi}{\delta\phi\delta\phi}\right)^{-1}\right\}.
\ee 
We stress here that, although the right-hand side of eq.(\ref{evolG}) has the structure of a one-loop 
correction, this evolution equation provides a ressumation of all order in $\hbar$, since the 
effective action appearing in the trace contains the dressed action, and not the bare one.
Eq.(\ref{evolG}) is therefore a self-consistent equation, in the spirit of a differential
Schwinger-Dyson equation, and is thus non perturbative.\\
In order to extract the evolution of the dressed parameters defining the quantum theory, though, we need to
adopt an approximation scheme and   
we assume, in the framework of the gradient expansion, the local potential 
approximation for $\Gamma$, with the kinetic term frozen to its classical expression, such that
\be\label{parametrization}
\Gamma_\xi=\int dt dz\Bigg\{\hf\partial_\mu\phi\partial^\mu\phi-U_\xi(\phi)
+\left[1-\tanh^2(\zeta)\right]V_\xi(\phi)-\tanh(\zeta)Y_\xi(\phi)\Bigg\},
\ee
where $U_\xi,V_\xi,Y_\xi$ are dressed potentials which define the quantum theory living on the kink,
and depend on the parameter $\xi$.
These potentials will be determined by plugging the ansatz (\ref{parametrization}) into the evolution 
equation (\ref{evolG}), and they read, at the tree-level,
\bea\label{tree_level}
U^{tree}(\phi)&=&\xi m_0^2\phi^2+\frac{\lambda_0}{24}\phi^4\nn
V^{tree}(\phi)&=&\frac{3}{2}m_0^2\phi^2\nn
Y^{tree}(\phi)&=&\frac{g_0}{6}\phi^3.
\eea
In order to respect the symmetries of the bare action, in what follows we consider even potentials
$U_\xi,V_\xi$ and an odd potential $Y_\xi$

\section{Evolution of the dressed parameters}

In order to derive the evolution of the dressed parameters, we 
have to compute the trace appearing in the evolution equation (\ref{evolG}), for a given configuration $\phi$.
Because of the symmetry of the function $\tanh(\zeta)$, a constant configuration for $\phi$ is not
appropriate, as in such a case the derivative $\dot Y_\xi$ does not appear in the left hand side of eq.(\ref{evolG}).
The appropriate choice here is the step-like configuration
\be\label{step1}
\phi_{step}=\mbox{sign}(z)\phi_0,
\ee
where $\phi_0$ is a constant. This configuration has a singular kinetic term, but the corresponding
singularity is $\xi$-independent in the framework of the gradient expansion (\ref{parametrization}),
and therefore has no influence on the evolution in $\xi$.
With such a configuration, the left hand side of the evolution equation (\ref{evolG}) is
\be
LT\left[m_0^2\phi_0^2-\dot U_\xi(\phi_0)-\dot Y_\xi(\phi_0)\right]
+\frac{T}{m_0}\left[2\dot V_\xi(\phi_0)+\ln 2 ~\dot Y_\xi(\phi_0)\right],
\ee
where $T$ is the length of the time dimension and $L$ is the length of the space dimension. 
These lengths being independent, one can independently identify in eq.(\ref{evolG})
the terms proportional to $T$ and the terms proportional to $LT$.

The second derivative of the effective action is, for the configuration (\ref{step1}),
\bea\label{2nd}
\frac{\delta^2\Gamma_\xi}{\delta\phi_1\delta\phi_2}&=&
-\left\{\partial_\mu\partial^\mu+U^{''}_\xi(\phi_0)\right\}\delta(t_1-t_2)\delta(z_1-z_2)\\
&&+\left\{\left[1-\tanh^2(\zeta)\right]V^{''}_\xi(\phi_0)-|\tanh(\zeta)|Y^{''}_\xi(\phi_0)\right\}
\delta(t_1-t_2)\delta(z_1-z_2).\nonumber
\eea
We need then the Fourier transform of the functions $|\tanh(\zeta)|$ and $1-\tanh^2(\zeta)$, and we find in Appendix A 
\bea
\int_{-\infty}^\infty dz~e^{-ikz}\left[1-\tanh^2(\zeta)\right]&\simeq&
4\frac{m_0^2}{k^3}\sin\left(\frac{k}{m_0}\right)-4\frac{m_0}{k^2}\cos\left(\frac{k}{m_0}\right)\nn
\int_{-\infty}^\infty dz~e^{-ikz}|\tanh(\zeta)|
&\simeq&2\pi\delta(k)+2\frac{m_0}{k^2}\left[\cos\left(\frac{k}{m_0}\right)-1\right].
\eea
We are interested in the limit of a strongly localized topological defect, and therefore consider 
the first order in $1/m_0$ only, where the previous Fourier transforms are
\bea
\int_{-\infty}^\infty dz~e^{-ikz}\left[1-\tanh^2(\zeta)\right]
&\simeq&\frac{4}{3m_0}\nn
\int_{-\infty}^\infty dz~e^{-ikz}|\tanh(\zeta)|
&\simeq&2\pi\delta(k)-\frac{1}{m_0}.
\eea
The Fourier transform of the second functional derivative (\ref{2nd}) is then
\bea\label{2ndF}
\frac{\delta^2\Gamma_\xi}{\delta\phi_1\delta\phi_2}&\simeq&
\left\{\omega_1^2-k_1^2-U_\xi^{''}(\phi_0)-Y_\xi^{''}(\phi_0)\right\}2\pi\delta(\omega_1+\omega_2)2\pi\delta(k_1+k_2)\\
&&+\frac{1}{3m_0}\left\lbrace 4V_\xi^{''}(\phi_0)+3Y_\xi^{''}(\phi_0)\right\}2\pi\delta(\omega_1+\omega_2)\nonumber
\eea
where we observe that, since translation invariance is broken in the space dimension,
there is no conservation of momentum $k$ in this direction.
In what follows, we give the main steps of the derivations only, and the details can be found in Appendix A.

\subsection{Evolution of the potentials}

For the step-like configuration (\ref{step1}), we evaluate the 
inverse of the second derivative (\ref{2ndF}) using the expansion
\be\label{ABA}
(A+B)^{-1}=A^{-1}-A^{-1}BA^{-1}+A^{-1}BA^{-1}BA^{-1}+\cdot\cdot\cdot,
\ee
where $A$ is proportional to $\delta(\omega_1+\omega_2)\delta(k_1+k_2)$ and thus is diagonal, 
and $B$ is proportional to $\delta(\omega_1+\omega_2)$ only and thus is off-diagonal in the 
space dimension. In the previous expansion, the small parameter is $k/m_0$, where $k$ is a typical IR momentum.
The identification of the terms proportional to $LT$ in the trace of eq.(\ref{evolG}) gives then:
\be\label{evolUeffL}
\dot U_\xi(\phi_0)+\dot Y_\xi(\phi_0)=m_0^2\phi_0^2
+\frac{m_0^2}{4\pi}\ln\left( 1+\frac{\Lambda^2}{U_\xi^{''}(\phi_0)+Y^{''}_\xi(\phi_0)}\right),
\ee
where a prime denotes a derivative with respect to the constant configuration $\phi_0$,
and $\Lambda$ is the UV cut off.
The latter will actually not appear in the evolution equations
for the parameters, since the expansion of eq.(\ref{evolUeffL}) in powers of $\phi_0$ leads to a 
field-independent divergence. We choose then the potentials such that $U_\xi(0)=0,Y_\xi(0)=0$, and 
substract the corresponding evolution equation from eq.(\ref{evolUeffL}) to obtain,
in the limit $\Lambda\to\infty$,
\be\label{evolU}
\dot U_\xi(\phi_0)+\dot Y_\xi(\phi_0)=m_0^2\phi_0^2
+\frac{m_0^2}{4\pi}\ln\left(\frac{U_\xi^{''}(0)+Y^{''}_\xi(0)}{U_\xi^{''}(\phi_0)+Y^{''}_\xi(\phi_0)}\right). 
\ee
The cut off does not appear in our evolution equation as a consequence of the derivative with respect
to a bare mass term, whereas we could expect logarithmic divergences in a 1+1 dimensional field theory. 
The projection of the equation (\ref{evolU}) on the subspace of 
even functions of $\phi_0$ gives the evolution of $U_\xi$, and its projection on the subspace
of odd functions gives the evolution of $Y_\xi$.

The evolution equations obtained after identification of the terms proportional to $T$ is
\be\label{evolV}
\dot V_\xi(\phi_0)+\frac{\ln 2}{2}~\dot Y_\xi(\phi_0)
=-\frac{m_0^2}{24\pi}\left( \frac{4V_\xi^{''}(\phi_0)+3Y_\xi^{''}(\phi_0)}{U_\xi^{''}(\phi_0)+Y_\xi^{''}(\phi_0)}
-\frac{4V_\xi^{''}(0)+3Y_\xi^{''}(0)}{U_\xi^{''}(0)+Y_\xi^{''}(0)}\right),
\ee
where the constant term has been chosen so as to respect $V_\xi(0)=0$. In this latter equation also,
the projection on the subspace of 
even functions gives the evolution of $V_\xi$, and its projection on the subspace
of odd functions gives the evolution of $Y_\xi$.

As is clear from eqs.(\ref{evolU},\ref{evolV}), a consistent solution for the potentials can be found 
only if $Y_\xi=0$: these two evolution equations cannot give identical evolutions for $Y_\xi$.
As a consequence, no odd function of the field appears in the effective theory. This property will be discussed in the 
last section, where we show that it is a consequence of symmetries of the quantum theory.

Finally, the effective action is 
\be\label{finalGamma}
\Gamma_\xi=\int dtdz\Bigg\{\hf\partial_\mu\phi\partial^\mu\phi-U_\xi(\phi)
+\left[1-\tanh^2(\zeta)\right]V_\xi(\phi)\Bigg\},
\ee
where the dressed potentials $U_\xi$ and $V_\xi$ satisfy the evolution equations
\bea\label{evolUV}
\dot U_\xi(\phi_0)&=&m_0^2\phi_0^2
+\frac{m_0^2}{4\pi}\ln\left(\frac{U_\xi^{''}(0)}{U_\xi^{''}(\phi_0)}\right)\\
\dot V_\xi(\phi_0)&=&-\frac{m_0^2}{6\pi}\left(\frac{V_\xi^{''}(\phi_0)}{U_\xi^{''}(\phi_0)}
-\frac{V_\xi^{''}(0)}{U_\xi^{''}(0)}\right).\nonumber
\eea
We observe that, in the framework of the gradient expansion (\ref{parametrization}), 
the evolution equation for $U_\xi$ is independent of $V_\xi$. A further step in the gradient expansion 
would consist in taking into account quantum fluctuations in the kinetic term, and 
write a general operator of the form $Z_\xi(\phi)\partial_\mu\phi\partial^\mu\phi$ in the effective action. 
The function $Z_\xi$ would then couple the evolution equations for $U_\xi$ and $V_\xi$.

Finally, we note that taking into account additional terms in the expansion (\ref{ABA}) would not
influence the evolution of the effective potential $U_\xi$, but would add corrections of higher orders
in $1/m_0$ to the evolution of $V_\xi$.

\subsection{Truncation of the dressed potentials}

Quantum fluctuations generate all the powers of field in the dressed potentials $U_\xi,V_\xi$. As discussed already,
no operator is irrelevant here, in the Wilsonian sense, and therefore in principle one should 
take into account all the powers of $\phi$. But if we assume small quantum fluctuations, 
we consider then the following truncation of the dressed potentials:
\bea
U_\xi(\phi_0)&=&\frac{M^2}{2}\phi_0^2+\frac{\lambda}{24}\phi_0^4+\frac{\beta}{6!}\phi_0^6\nn
V_\xi(\phi_0)&=&\frac{v_1}{2}\phi_0^2+\frac{v_2}{24}\phi_0^4+\frac{v_3}{6!}\phi_0^6,
\eea
where the parameters $M^2,\lambda,\beta,v_1,v_2,v_3$ depend on $\xi$.
This truncation takes into account the interactions which appear in the bare theory, as well as the lowest interaction ($\phi_0^6$) generated by quantum fluctuations.
An expansion in powers of $\phi_0$ in the evolution equation (\ref{evolUV}) for $U_\xi$ gives, after identification
of the different powers,
\bea\label{sol1}
&&\mbox{order}~\phi_0^2:~~~~~~~~M\dot M=m_0^2-\frac{\lambda m_0^2}{8\pi M^2}\nn
&&\mbox{order}~\phi_0^4:~~~~~~~~\dot\lambda=\frac{3m_0^2}{4\pi M^2}\left( \frac{\lambda^2}{M^2}-\frac{\beta}{3}\right) \nn
&&\mbox{order}~\phi_0^6:~~~~~~~~\dot\beta=\frac{15\lambda m_0^2}{2\pi M^4}\left( \frac{\beta}{2}-\frac{\lambda^2}{M^2}\right)
\eea
The identification of the powers of $\phi_0$ in the evolution equation (\ref{evolUV}) for $V_\xi$ gives
\bea\label{sol2}
&&\mbox{order}~\phi_0^2:~~~~~~~~\dot v_1=\frac{m_0^2}{6\pi M^2}\left( \frac{\lambda v_1}{M^2}-v_2\right) \\
&&\mbox{order}~\phi_0^4:~~~~~~~~\dot v_2=-\frac{m_0^2}{\pi M^2}\left( \frac{\lambda^2 v_1}{M^4}
-\frac{\beta v_1}{6M^2}-\frac{\lambda v_2}{M^2}+\frac{v_3}{6}\right) \nn
&&\mbox{order}~\phi_0^6:~~~~~~~~\dot v_3=-\frac{15m_0^2}{\pi M^4}\left[ \frac{\lambda v_1}{M^2}
\left( \frac{\beta}{2}-\frac{\lambda^2}{M^2}\right) +v_2\left( \frac{\lambda^2}{M^2}-\frac{\beta}{6}\right) 
-\frac{\lambda v_3}{6}\right]. \nonumber
\eea
If one desires to obtain the evolution of the dressed parameters with $\xi$, it is possible to
solve the equations (\ref{sol1},\ref{sol2}) numerically, but we give in what follows approximate analytical solutions,
which contain the essential properties of the quantum theory.

\subsection{One loop approximation}

We study here the one-loop approximation of the non-perturbative evolution equations for $M,\lambda,\beta$.
For this, we note that the right hand side of eq.(\ref{evolG}) contains the
quantum corrections, such that the one-loop approximation 
is obtained by replacing on the right hand side the dressed parameters by the bare ones: $M\to\sqrt{2\xi}m_0$, 
$\lambda\to\lambda_0$ and $\beta\to 0$. We obtain then
for the one-loop parameters $M^{(1)},\lambda^{(1)},\beta^{(1)}$
\bea\label{oneloop}
M^{(1)}\dot M^{(1)}&=&m_0^2-\frac{\lambda_0}{16\pi\xi}\nn
\dot\lambda^{(1)}&=&\frac{3\lambda_0^2}{16\pi\xi^2 m_0^2}\nn
\dot\beta^{(1)}&=&-\frac{15\lambda_0^3}{16\pi\xi^3 m_0^4}.
\eea
It is interesting to compare these results with usual Feynman diagrams, obtained from the bare theory
\be
\int dtdz\left\lbrace \frac{1}{2}\partial_\mu\phi\partial^\mu\phi-\xi m_0^2\phi^2-\frac{\lambda_0}{24}\phi^4\right\rbrace ,
\ee 
i.e. the initial bare theory without the $z$-dependent quadratic and cubic terms.

The one-loop correction to the parameter $M^2$ is generated by the interaction
$\phi^4$ which is represented by the tadpole diagram
\bea
(M^{(1)})^2-2\xi m_0^2
&=&\frac{i\lambda_0}{2}\int\frac{d^2p}{(2\pi)^2}\frac{1}{p^2-2\xi m_0^2}\\
&=&\frac{\lambda_0}{2}\frac{\Omega_2}{(2\pi)^2}\int_0^\Lambda \frac{qdq}{q^2+2\xi m_0^2}\nn
&=&\frac{\lambda_0}{8\pi}\ln\left( 1+\frac{\Lambda^2}{2\xi m_0^2}\right),\nonumber
\eea
where the factor $1/2$ takes into account the symmetry factor of the graph.
It can be checked that the derivative of the latter result with respect to $\xi$ indeed gives the
expected expression (\ref{oneloop}) for $\partial_\xi[(M^{(1)})^2-2\xi m_0^2]=2(M\dot M^{(1)}-m_0^2)$, in the limit 
$\Lambda\to\infty$.

The one-loop correction to the coupling $\lambda$ is given by
\bea
\lambda^{(1)}-\lambda_0&=&\frac{3i\lambda_0^2}{2}\int\frac{d^2p}{(2\pi)^2}\frac{1}{(p^2-2\xi m_0^2)^2}\\
&=&-\frac{3\lambda_0^2}{2}\frac{\Omega_2}{(2\pi)^2}\int_0^\infty \frac{qdq}{(q^2+2\xi m_0^2)^2}\nn
&=&-\frac{3\lambda_0^2}{16\pi\xi m_0^2},\nonumber
\eea
where $3\lambda_0/2$ in the first line takes into account the symmetry factor and the 3 permutations
of vanishing incoming momenta.
The derivative of the latter result with respect to $\xi$ indeed gives the above expression
(\ref{oneloop}) for $\dot\lambda^{(1)}$.

The coupling $\beta$ is generated by quantum fluctuations, and its one-loop expression is, taking into 
account the symmetry and permutation factors,
\bea
\beta^{(1)}&=&\frac{6!}{8\times 6}(i\lambda_0)^3\int\frac{d^2p}{(2\pi)^2}\frac{1}{(p^2-2\xi m_0^2)^3}\\
&=&15\lambda_0^3\frac{\Omega_2}{(2\pi)^2}\int_0^\infty \frac{qdq}{(q^2+2\xi m_0^2)^3}\nn
&=&\frac{15\lambda_0^3}{32\pi\xi^2 m_0^4},\nonumber
\eea
The derivative of the latter result with respect to $\xi$ indeed gives the above expression
(\ref{oneloop}) for $\dot\beta^{(1)}$.

We checked here that our non perturbative evolution equations (\ref{sol1}) are consistent, at one loop,
with usual Feynman graphs. This feature is a consequence of the fact that, in the framework of the
gradient expansion (\ref{parametrization}), the evolution of $U_\xi$ is independent of the 
evolution of $V_\xi$. Beyond one loop, the gradient expansion does not give the same results than
the loop expansion, as it is based on an expansion in powers of the momentum.

\subsection{Approximate analytical solution}

We are interested here in approximate analytical solutions for the parameters $\lambda$ and $v_1$.

An approximate solution for $\lambda$ given in eqs.(\ref{sol1}) can be obtained by keeping the bare values
for the other parameters: $M\to\sqrt{2\xi}m_0$ and $\beta\to 0$, in which case the equation for $\lambda$ reads
\be\label{approxlm1}
\frac{\dot\lambda}{\lambda^2}=\frac{3}{16\pi\xi^2 m_0^2}
\ee
We see that, as quantum fluctuations arise ($\xi$ decreases),
$\lambda$ decreases ($\dot\lambda>0$), which was expected, as a scalar self coupling is known to decrease
in the IR. If we define the renormalized coupling $\lambda_R=\lambda(1)$, the solution 
of eq.(\ref{approxlm1}) can easily be found and reads:
\be\label{lambdaapprox}
\lambda(\xi)=\lambda_R\left[1+\frac{3\lambda_R}{16\pi m_0^2}\left( \frac{1}{\xi}-1\right) \right]^{-1}.
\ee
In the spirit of the present functional method, 
the bare coupling $\lambda_0$ should be found in the limit $\xi\to\infty$, which leads to the following expression
for the dressed coupling in terms of the bare coupling
\be\label{lambdaR}
\lambda_R=\lambda_0\left( 1+\frac{3\lambda_0}{16\pi m_0^2}\right)^{-1}.
\ee
Using a similar approximation in the evolution equation for the parameter $v_1$,
i.e. $M\to\sqrt{2\xi}m_0$ and $v_2\to 0$, we find the following evolution for $v_1$
\be 
\frac{\dot v_1}{v_1}=\frac{\lambda}{24\pi\xi^2m_0^2},
\ee
where $\lambda$ is given by eq.(\ref{lambdaapprox}). The integration of this equation gives then
\be 
v_1(\xi)=3m_R^2\left[1+\frac{3\lambda_R}{16\pi m_0^2}\left(\frac{1}{\xi}-1\right)\right]^{-2/9},
\ee 
where we define $3m_R^2=v_1(1)$. As previously, a relation between the renormalized parameters
$\lambda_R$ and $m_R^2$ can be obtained, by taking the limit $\xi\to\infty$ in the previous equation, 
with $v_1\to 3m_0^2$, such that
\be
m_R^2=m_0^2\left(1-\frac{3\lambda_R}{16\pi m_0^2}\right)^{2/9}.
\ee 
From the relation (\ref{lambdaR}), we obtain then the following expression for $m_R^2$ in terms of the bare parameters only
\be\label{mR}
m_R^2=m_0^2\left(\frac{16\pi m_0^2}{3\lambda_0+16\pi m_0^2}\right)^{2/9}.
\ee 
Eqs.(\ref{lambdaR},\ref{mR}) consist in a ressumations in all orders in $\hbar$, and are derived in the
limit of highly localized topological defect, $m_0^2>>\lambda_0$.

\section{Discussion}

We now discuss the vanishing of the dressed term $\tanh(\zeta)Y_\xi(\phi)$ in the effective action, 
as a consequence of a discrete symmetry of the theory.

In this work, we considered the quantization of fluctuations above the kink $\Phi_{bg}$ given in eq.(\ref{kink}), but
the classical equation of motion (\ref{eq.mot.}) has actually two kink solutions centered on $z=0$, which 
are $\pm\Phi_{bg}$. We now discuss the symmetry of the quantum theory under the transformation 
$\Phi_{bg}\to -\Phi_{bg}$, and we denote by an upper indice $^{(\pm)}$ the different quantities 
defined respectively on the backgrounds $\pm\Phi_{bg}$.
We note here that the vacuum of the theory, the constant configuration $\Phi_0=m_0\sqrt{6/\lambda_0}$, does not
respect the symmetry $\Phi_0\to -\Phi_0$, since this symmetry is spontaneously broken. 

The bare action corresponding to the background $-\Phi_{bg}$ is
\bea
S_\xi^{(-)}[\phi]&=&\int dt dz\Bigg\{\hf\partial_\mu\tilde\Phi\partial^\mu\tilde\Phi
-\xi m_0^2\tilde\Phi^2-\frac{\lambda_0}{24}\tilde\Phi^4\nn
&&~~~~~~~~~~~+\frac{3}{2}m_0^2\left[1-\tanh^2(\zeta)\right]\tilde\Phi^2
+\frac{g_0}{6}\tanh(\zeta)\tilde\Phi^3\Bigg\}\nn
&=&S_\xi^{(+)}[\psi],
\eea
where $\psi(t,z)=\phi(t,-z)$. The source term can then be written
\be
\int dt dz ~j\phi=\int dt dz ~g\psi,
\ee
where $g(t,z)=j(t,-z)$, such that the partition function is
\bea
Z_\xi^{(-)}[j]&=&\int{\cal D}[\phi]\exp\left\{iS^{(-)}[\phi]+i\int dtdz~j\phi\right\}\nn
&=&\int{\cal D}[\psi]\exp\left\{iS^{(+)}[\psi]+i\int dtdz~g\psi\right\}\nn
&=&Z_\xi^{(+)}[g].
\eea
The classical field corresponding to the background $-\Phi_{bg}$ is
\bea
\phi_c^{(-)}(t,z)&=&\frac{\delta W_\xi^{(-)}}{\delta j(t,z)}
=\int dsdy\frac{\delta W_\xi^{(+)}}{\delta g(s,y)}\frac{\delta g(s,y)}{\delta j(t,z)}\nn
&=&\int dsdy\frac{\delta W_\xi^{(+)}}{\delta g(s,y)}\delta(y+z)\delta(s-t)
=\frac{\delta W_\xi^{(+)}}{\delta g(t,-z)}
=\frac{\delta W_\xi^{(+)}}{\delta j(t,z)}\nn
&=&\phi_c^{(+)}(t,z).\nonumber
\eea
and is therefore independent of the sign of the background: $\phi_c^{(-)}=\phi_c^{(+)}$.
When defining the Legendre transform $\Gamma_\xi$, we inverse the relation $j\to\phi_c$, such that the source
is now function of the background, and therefore $j^{(-)}=j^{(+)}$. The effective action is then
\bea
\Gamma_\xi^{(-)}[\phi_c]&=&W_\xi[j^{(-)}]-\int dtdz~j^{(-)}\phi_c\nn
&=&W_\xi[j^{(+)}]-\int dt dz~j^{(+)}\phi_c\nn
&=&\Gamma_\xi^{(+)}[\phi_c].
\eea
As a consequence, the effective action does not depend on the sign of the background, such that the 
dressed term $\tanh(\zeta)Y_\xi(\phi_c)$ in the effective action (\ref{parametrization})
must vanish, as it should satisfy $-Y_\xi=Y_\xi$.
The corresponding term in the bare action does not survive quantization.
It is interesting to note that the non-perturbative method presented here allows us to see the vanishing
of the dressed potential $Y_\xi$, using eqs.(\ref{evolU},\ref{evolV}), which
 means that quantum fluctuations are strong enough to 
cancel the corresponding term present in the bare action. This could not be obtained 
within a perturbative approach, but only within a method using a self consistent equation as eq.(\ref{evolG}).

A next step in this study consists in including fermions coupled to the scalar field fluctuating
over the background kink. From the results obtained here, we can expect a usual Yukawa coupling $\phi\ol\psi\psi$ to 
be relevant to the problem, or more generally a coupling of the form $f(\phi)\ol\psi\psi$,
without an explicit $z$-dependence. As explained in Appendix B, the 
evolution equation for the effective action $\Gamma_\xi$ with $\xi$ is then obtained in the same way,
with more involved calculations though, as the second derivative $\delta^{(2)}\Gamma$ is then a 
$3\times 3$ matrix, with rows $\phi,\ol\psi,\psi$, and the computation of the trace in eq.(\ref{evolG})
involves the inverse of this matrix.
In higher dimensions, the method presented here is of course valid, and can include any other matter field. 
Also, it can be extended to higher symmetries and deal with gauge fields. As far as 
supersymmetry is concerned, 
the use of superfield formalism necessitates a modification of the evolution equation (\ref{evolG}),
which takes into account the chirality constraint of the superfields.

Finally, we emphasize the advantage of the present approach, in 1+1 dimensions, compared to the
Wilsonian approach: we were able to generate non perturbative flows without referring to a running cut off,
as no cut off appears in the evolution for the dressed potentials, and the integration 
of the corresponding flows led us to cut-off-free relations between bare and dressed parameters.

\vspace{1cm}

\nin{\bf Acknowledgments:}
This work is supported by the EPEAEK program "Pythagoras", and co-funded by
the European Union (75\%) and the Hellenic State (25\%).

\section*{Appendix A: Computation of the trace}

For the step-like configuration $\phi=\mbox{sign}(z)\phi_0$, the second derivative of the effective action is
\bea
\frac{\delta^2\Gamma_\xi}{\delta\phi_1\delta\phi_2}&=&\Big\{-\partial_\mu\partial^\mu-U_\xi^{''}(\phi_0)
+\left[1-\tanh^2(\zeta)\right]V_\xi^{''}(\phi_0)\nn
&&~~-|\tanh(\zeta)|Y_\xi^{''}(\phi_0)\Big\}\delta(t_1-t_2)\delta(z_1-z_2),
\eea
and therefore we need the Fourier transform of $1-\tanh^2(\zeta)$ and $|\tanh(\zeta)|$. 
For this, we approximate the function $\tanh(\zeta)$ by $\zeta$ in the interval $[-1;1]$ and by 0 
outside this interval. This approximation captures the essential features of the kink, 
and leads, for the Fourier transform of $1-\tanh^2(\zeta)$, to
\bea
&&\int_{-\infty}^\infty dz~e^{-ikz}\left[1-\tanh^2(\zeta)\right]\nn
&\simeq&\int_{-1/m_0}^{1/m_0} dz~e^{-ikz}\left(1-\zeta^2\right)\\
&=&4\frac{m_0^2}{k^3}\sin\left(\frac{k}{m_0}\right)
-4\frac{m_0}{k^2}\cos\left(\frac{k}{m_0}\right)\nonumber.
\eea
The same approximation leads, for the Fourier transform of $|\tanh(\zeta)|$, to
\bea
&&\int_{-\infty}^\infty dz~e^{-ikz}|\tanh(\zeta)|\\
&\simeq&\int_{-\infty}^\infty dz~e^{-ikz}+\int_{-1/m_0}^{1/m_0} dz e^{-ikz}\left(|\zeta|-1\right)\nn
&=&2\pi\delta(k)+2\frac{m_0}{k^2}\left[\cos\left(\frac{k}{m_0}\right)-1\right]\nonumber
\eea
Since we are interested in the limit of a highly localized topological defect, we
consider the situation where $m_0>>|k|$, which gives
\bea
\int_{-\infty}^\infty dz~e^{-ikz}\left[1-\tanh^2(\zeta)\right]
&\simeq&\frac{4}{3m_0}+{\cal O}\left(\frac{k^2}{m_0^3}\right)\nn
\int_{-\infty}^\infty dz~e^{-ikz}|\tanh(\zeta)|
&\simeq&2\pi\delta(k)-\frac{1}{m_0}+{\cal O}\left(\frac{k^2}{m_0^3}\right)\nonumber
\eea
The inverse $(\delta^2\Gamma)^{-1}$ is taken as
\be\label{exp}
(A+B)^{-1}=A^{-1}-A^{-1}BA^{-1}+A^{-1}BA^{-1}BA^{-1}+\cdot\cdot\cdot,
\ee
where $A$ is proportional to $\delta(\omega_1+\omega_2)\delta(k_1+k_2)$ and thus is diagonal, 
and $B$ is proportional to $\delta(\omega_1+\omega_2)$ only and thus is off-diagonal in the 
space dimension.
We obtain then, taking into account the first order in $B$ in the expansion (\ref{exp})
\bea
\left(\frac{\delta^2\Gamma_\xi}{\delta\phi_1\delta\phi_2}\right)^{-1}&\simeq&
\frac{2\pi\delta(\omega_1+\omega_2)2\pi\delta(k_1+k_2)}{\omega_1^2-k_1^2-U_\xi^{''}(\phi_0)-Y_\xi^{''}(\phi_0)}\\
&&-\frac{(3m_0)^{-1}\left[4V_\xi^{''}(\phi_0)+3Y_\xi^{''}(\phi_0)\right]2\pi\delta(\omega_1+\omega_2)}
{\left[\omega_1^2-k_1^2-U_\xi^{''}(\phi_0)-Y_\xi^{''}(\phi_0)\right]
\left[\omega_2^2-k_2^2-U_\xi^{''}(\phi_0)-Y_\xi^{''}(\phi_0)\right]}.\nonumber
\eea
The term proportional to $LT$ in the trace of eq.(\ref{evolG}) is
\bea\label{LT}
&&LT\int\frac{d\omega}{2\pi}\frac{dk}{2\pi}\frac{1}{\omega^2-k^2-U_\xi^{''}(\phi_0)- Y_\xi^{''}(\phi_0)}\nn
&=&-iLT\frac{\Omega_2}{(2\pi)^2}\int_0^\Lambda \frac{qdq}{q^2+U_\xi^{''}(\phi_0)+Y_\xi^{''}(\phi_0)}\nn
&=&-LT\frac{i}{4\pi}\ln\left( 1+\frac{\Lambda^2}{U_\xi^{''}(\phi_0)+Y_\xi^{''}(\phi_0)}\right) .
\eea
where $q$ is the Euclidean 2-momentum and $\Omega_2=2\pi$ is the solid angle in dimension 2.\\
The term proportional to $T$ only in the trace of eq.(\ref{evolG}) is
\bea\label{Tonly}
&&-\frac{4V^{''}_\xi(\phi_0)+3Y^{''}_\xi(\phi_0)}{3m_0}T
\int\frac{d\omega}{2\pi}\frac{dk}{2\pi}\frac{1}{\left(\omega^2-k^2-U_\xi^{''}(\phi_0)-Y_\xi^{''}(\phi_0)\right)^2}\nn
&=&-\frac{4V^{''}_\xi(\phi_0)+3Y^{''}_\xi(\phi_0)}{3m_0}T
\frac{\Omega_2}{(2\pi)^2}\int_0^\infty\frac{qdq}{\left(q^2+U_\xi^{''}(\phi_0)+Y_\xi^{''}(\phi_0)\right)^2}\nn
&=&\frac{-iT}{12\pi m_0}
\frac{4V^{''}_\xi(\phi_0)+3Y_\xi^{''}(\phi_0)}{U_\xi^{''}(\phi_0)+Y_\xi^{''}(\phi_0)}\nn
\eea
The left-hand side of eq.(\ref{evolG}) is, for the step-like configuration,
\be\label{lefthand}
LT\left[m_0^2\phi_0^2-\dot U_\xi(\phi_0)-\dot Y_\xi(\phi_0)\right]
+\frac{T}{m_0}\left[2\dot V_\xi(\phi_0)+\ln 2 \dot Y_\xi(\phi_0)\right],
\ee
and, together with eqs.(\ref{LT},\ref{Tonly}), we obtain for the evolution of the dressed potentials
\bea
\dot U_\xi(\phi_0)+\dot Y_\xi(\phi_0)&=&m_0^2\phi_0^2
+\frac{m_0^2}{4\pi}\ln\left( 1+\frac{\Lambda^2}{U_\xi^{''}(\phi_0)+Y_\xi^{''}(\phi_0)}\right)\\
\dot V_\xi(\phi_0)+\frac{\ln 2}{2}~\dot Y_\xi(\phi_0)
&=&-\frac{m_0^2}{24\pi}\frac{4V_\xi^{''}(\phi_0)+3Y_\xi^{''}(\phi_0)}{U_\xi^{''}(\phi_0)+Y_\xi^{''}(\phi_0)}
.\nonumber
\eea

\section*{Appendix B: Extension to a Yukawa interaction}

We give here the main steps of the extension of the previous method to a Yukawa interaction
in $d+1$ dimensions, where the kink expands in the extra dimension, with coordinate $z$. For $d\ge 4$,
the theory is not renormalizable, and we consider it an effective theory, valid up to 
an energy scale $\Lambda$, which is our cut off.\\
The bare action is, for massless fermions,
\be
S_0=\int d^dx dz\left\{i\ol\Psi\br\partial\Psi+\hf\partial_\mu\Phi\partial^\mu\Phi
-\eta_0\Phi\ol\Psi\Psi-U_B(\Phi)\right\},
\ee
where the scalar potential $U_B(\Phi)$ is given in eq.(\ref{barepot}). The fermion 
field having no expectation value, the kink configuration is the same as in eq.(\ref{kink}), and the action 
to quantize is
\bea
S_\xi&=&\int d^dx dz\Bigg\{i\ol\Psi\br\partial\Psi
-\eta_0\Phi_{bg}(\zeta)\ol\Psi\Psi -\eta_0\tilde\Phi\ol\Psi\Psi\nn
&&~~~~~~~~~~+\hf\partial_\mu\tilde\Phi\partial^\mu\tilde\Phi
-\xi m_0^2\tilde\Phi^2-\frac{\lambda_0}{24}\tilde\Phi^4\nn
&&~~~~~~~~~~~+\frac{3}{2}m_0^2\left[1-\tanh^2(\zeta)\right]\tilde\Phi^2
-\frac{g_0}{6}\tanh(\zeta)\tilde\Phi^3\Bigg\},
\eea
where $\tilde\Phi$ represent the fluctuations above the classical kink. In the previous
expression, the $z$-dependent mass term $\eta_0\Phi_{bg}(\zeta)\ol\Psi\Psi$ for the fermion is responsible for the fermion
localization on the brane $z=0$, as discussed in \cite{local}. \\
The partition function, functional of the sources $j,\eta,\ol\eta$, is
\bea
Z_\xi&=&\int{\cal D}[\tilde\Phi,\Psi,\ol\Psi]\exp\left\{ iS_\xi[\tilde\Phi,\Psi,\ol\Psi]
+i\int d^dxdz\left( j\tilde\Phi+\ol\eta\Psi+\ol\Psi\eta\right)\right\} \nn 
&=&\exp\left( iW_\xi[j,\eta,\ol\eta]\right) ,
\eea
from which the classical fields $(\phi,\psi,\ol\psi)$ are defined:
\bea
\frac{\delta W_\xi}{\delta j}&=&\phi\nn
\frac{\delta W_\xi}{\delta\ol\eta}&=&\psi\nn
\frac{\delta W_\xi}{\delta\eta}&=&-\ol\psi.
\eea
The proper graphs generator functional of the classical fields $\phi,\ol\psi,\psi$ 
is defined as the Legendre transform of $W$, after inverting the relations 
$(j,\eta,\ol\eta)\to(\phi_\xi,\psi_\xi,\ol\psi_\xi)$ to $(\phi,\psi,\ol\psi)\to(j_\xi,\eta_\xi,\ol\eta_\xi)$:
\be
\Gamma_\xi[\phi,\psi,\ol\psi]=W_\xi[j,\eta,\ol\eta]-\int d^dx dz \left(j_\xi\phi+\ol\eta_\xi\psi+\ol\psi\eta_\xi\right), 
\ee
and its functional derivatives are
\bea
\frac{\delta\Gamma_\xi}{\delta\phi}&=&-j_\xi\nn
\frac{\delta\Gamma_\xi}{\delta\ol\psi}&=&-\eta_\xi\nn
\frac{\delta\Gamma_\xi}{\delta\psi}&=&\ol\eta_\xi.
\eea
The evolution equation for $\Gamma_\xi$ with $\xi$ is derived in the same way as was done in 1+1 dimensions and reads:
\be\label{evolGbis}
\dot\Gamma_\xi+m_0^2\int d^dx dz~\phi^2
=-im_0^2~\mbox{Tr}\left\{\left(\delta^2\Gamma_\xi\right)_{\phi\phi}^{-1}\right\},
\ee 
but this time $(\delta^2\Gamma_\xi)^{-1}_{\phi\phi}$ is the $\phi\phi$ component of the inverse of the matrix
\be
\delta^2\Gamma_\xi=\left( \begin{array}{ccc}
\frac{\delta^2\Gamma}{\delta\ol\psi\delta\psi}
&\frac{\delta^2\Gamma}{\delta\ol\psi\delta\ol\psi}
&\frac{\delta^2\Gamma}{\delta\ol\psi\delta\phi}\\
\frac{\delta^2\Gamma}{\delta\psi\delta\psi}
&\frac{\delta^2\Gamma}{\delta\psi\delta\ol\psi}
&\frac{\delta^2\Gamma}{\delta\psi\delta\phi}\\
\frac{\delta^2\Gamma}{\delta\phi\delta\psi}
&\frac{\delta^2\Gamma}{\delta\phi\delta\ol\psi}
&\frac{\delta^2\Gamma}{\delta\phi\delta\phi}\\
                   \end{array}\right) .
\ee
Note that the components $\delta^2\Gamma_{\ol\psi\ol\psi}$ and $\delta^2\Gamma_{\psi\psi}$ do not
vanish in general, as quantum fluctuations generate four-fermion interactions and higher powers of $\ol\psi\psi$.
Also, compared to the 1+1 dimensional model, the trace in eq.(\ref{evolGbis}) contains divergences, such 
that the cut off $\Lambda$ will appear in the final equations.\\
In order to take into account fermion localization, and the symmetries of the system,
we propose here the following 
gradient expansion, where we disregard higher order fermion interactions and wave function renormalization, 
\bea
\Gamma_\xi&=&\int d^dx dz\Big\lbrace 
\hf\partial_\mu\phi\partial^\mu\phi-U^1_\xi(\phi)\ol\psi\psi-U^2_\xi(\phi)\nn
&&~~~~~+[1-\tanh^2(\zeta)]\left[i\ol\psi\br\partial\psi+V^1_\xi(\phi)\ol\psi\psi+V^2_\xi(\phi)\right] \Big\}.
\eea
In the previous expression, fermion localization is implemented via the $\zeta$-dependence fermion kinetic term: 
away from the brane, for $\zeta\ne0$, the fermion propagation is exponentially damped.
The consistency of this ansatz for the functional dependence of $\Gamma_\xi$ has to be checked when 
computing the trace in eq.(\ref{evolGbis}) and following the
evolution of $\Gamma_\xi$ with $\xi$, which leads to the evolution of the scalar potentials $U^{1,2}_\xi(\phi)$
and $V^{1,2}_\xi(\phi)$.\\
An interesting study is then to look for a possible mass generated dynamically, on the brane, for the would-be
massless fermion. This can be investigated using the present method, since it is based on a self consistent
equation, as was already done in the Kaluza-Klein framework \cite{kaluzaklein}. In the present context, 
the fermion dynamical mass, if there is one, is $m_{dyn}=U^1_\xi(0)-V^1_\xi(0)$.

\end{document}